\documentstyle[epsf]{mn}

\title{Effects of Gravity on the Structure of Postshock Accretion Flows in 
Magnetic Cataclysmic Variables}

\author[M. Cropper et al.]{ 
Mark Cropper$^{1}$, 
Kinwah Wu$^{2}$, 
Gavin Ramsay$^{1}$ and 
Aysegul Kocabiyik$^{1}$ \\ 
$^{1}$ Mullard Space Science Laboratory, University College London, Holmbury
St. Mary, Surrey, RH5~6NT.\\
$^{2}$ Research Centre for Theoretical Astrophysics, Department of Physics,
Sydney University, Sydney NSW, Australia.}

\date{Received: }

\begin{document}
\def\Mdot{\hbox{$\dot M$}}
\def\Msun{\hbox{$M_\odot$\,}}
\def\Rwd{\hbox{$R_{_{\rm WD}}$\,}}
\def\rchi{\hbox{$\chi_{red}^2$}\,}

\maketitle

\begin{abstract}
We have calculated the temperature and density structure of the hot
postshock plasma in magnetically confined accretion flows,
including the gravitational potential. This avoids the inconsistency
of previous calculations which assume that the height of the shock is
negligible. We assume a stratified accretion column with 1-d flow along  
the symmetry axis. 
We find that the calculations predict a lower shock temperature than
previous calculations, with a flatter temperature profile with height.
We have revised previous determinations of the masses of the white
dwarf primary stars and find that for higher mass white dwarfs 
there is a general reduction
in derived masses when the gravitational potential is included. This
is because the spectrum from such flows is harder than that of previous
prescriptions at intermediate energies.
\end{abstract}


\section{Introduction}

Magnetic Cataclysmic Variables (mCVs) are strong emitters of hard
X-rays. These are produced in the postshock region immediately above
the accreting white dwarf. In this postshock region there is a strong
temperature gradient between the shock front ($>10^{8}$ K) and the
surface of the white dwarf. By correctly modelling the properties of
the emission region it is possible to determine fundamental proporties
of the mCV, such as the mass of the white dwarf. For an overview of
the modelling the hard X-ray emission see Cropper et al. 1999, while
Cropper (1990) and Warner (1995) provide more general reviews of mCVs.

Wu (1994) (also Wu, Chanmugam \& Shaviv 1994) described a closed integral
solution for the temperature and density profile of this postshock flow. The
treatment included cyclotron cooling as well as thermal bremsstrahlung:
cyclotron cooling is important for those mCVs with strongly magnetic ($>10$ MG)
white dwarfs.  The advantage of this scheme is that the temperature and density
profiles can be calculated analytically (with the exception of a final
summation). Thus multi-temperature spectra can be calculated sufficiently
rapidly to permit spectral fits to X-ray data. Previously, an analytical
solution existed only for the pure bremsstrahlung case (Aizu 1973).
The assumptions for the boundary conditions in Wu (1994) are the same as those
in Aizu (1973): these are a cold preshock flow (at free-fall preshock 
velocities) followed by a strong shock and then
a hot postshock flow cooling onto a cold white dwarf
surface (reaching zero temperature and velocity at the base of
the flow). As in Aizu (1973), the height
of the shock was assumed to be negligible so that the effects due to gravity 
were neglected.

Cropper, Ramsay \& Wu (1998) and Ramsay et al. (1998) used the Wu (1994)
formulation to determine the masses of the white dwarfs in those mCVs observed
with {\it Ginga}. The masses they derived are at the higher end of
expectations, and in the case of XY Ari, are significantly higher than the best
determinations obtained by other means (Ramsay et al. 1998). This indicates 
that there
is more hard X-ray flux than expected, or less soft X-ray flux as a result of
significantly more complex forms of absorption than assumed in Cropper et
al. (1998). The latter has been explored for BY Cam in Done \& Magdziarz
(1998). The effect of absorption in mCVs is a topic of its own, and we do not
pursue it here. Issues of the emission from a {\it structured} 
accretion region can be dealt with to first order by the
summation of a sufficiently large number of local models with different
accretion rates and magnetic fields. This is because the emission from 
the postshock flow is optically thin in the continuum down to almost 
the white dwarf surface. Therefore, here we 
take the path of exploring the effect of reducing the restrictiveness of the
assumptions in the emission model itself. 

Several options present themselves at this stage. These include improvement of
the lower boundary condition to match the atmosphere of the white dwarf more
appropriately, separate treatment of the proton and electron populations and
more detailed treatment of the physics within the shock itself. It 
should be noted that there have been a number of calculations for the
postshock radial flows onto white dwarfs, for example Imamura \& Durisen
(1983) (and references therein) and Woelk \& Beuermann (1996). Some of these
include 2-fluid effects, Compton cooling and the effect of a gravitational 
potential. These are all
important. The aim of those calculations has been to determine the postshock
temperature and density, and predict spectra and X-ray light curves, but they
are unsuitable for the iterative model fitting of X-ray data. This differs from
the approach adopted by Wu (1994) and Cropper et al. (1998) which attempts to
extract information from the X-ray spectra and which requires a formulation
that can be computed sufficiently rapidly for that purpose. Here we continue
along the path of improvements to that technique
by addressing the elimination of a negligible shock height assumption. In so
doing we explore and elucidate clearly for the first time 
the effects of including a radially
varying gravitational acceleration on the postshock flow structure.

The justification for assuming that the shock height has negligible 
effect is
given, for example, in Frank, King \& Raine (1992). This is adequate for the
first generation of mass estimates as in Cropper et al. (1998): however, in
those cases where the cyclotron cooling is insignificant (the IPs) and where
the mass of the white dwarf is larger than $\sim0.8$\Msun, application of the 
Aizu (1973) or Wu
(1994) formulae results in very significant shock heights for typical specific
accretion rates (0.3 \Rwd for a 1.0 \Msun white dwarf accreting at 1.0 g
s$^{-1}$ cm$^{-2}$). This is inconsistent with the assumptions used by Aizu
(1973), Wu (1994) and Cropper et al. (1998) (where the gravitational
acceleration is ignored) and with Woelk \& Beuermann (1996) (who use a constant
gravitational acceleration). Since 
the temperature in the immediate postshock region 
is the pre-shock velocity at that height (divided by a factor 
4 because of the strong shock jump condition) this implies 
that the shock temperature will be significantly lower
than that calculated from the free fall velocity {\it at the surface} as used
when assuming a negligible shock height. For this paper, we have therefore
augmented the treatment in Wu (1994) to include the effect of the variation of
the gravitational acceleration within the postshock flow.

\section{Formulation}

We derive the following set of 1-d steady state conservation equations taking
into account the gravitational potential (see Appendix for details):
\begin{equation}
\frac{d}{dx}(\rho v) = 0
\end{equation}
\begin{equation}
\frac{d}{dx}(\rho v^{2} + P) = \frac{-GM\rho}{x^{2}}
\end{equation}
\begin{equation}
v\frac{dP}{dx} + \gamma P \frac{dv}{dx} = -(\gamma-1)\Lambda.
\end{equation}
With the ideal gas law, $P/\rho=kT/\mu m_{H}$, this set of equations is
closed. Here $x$ is the spatial coordinate, $\rho$ is the density, $v$ is the
flow velocity, $P$ is the pressure, $\gamma$ is the adiabatic index, $T$ is the
temperature, $\mu$ is the mean molecular mass and $m_{H}$ is the mass of a
hydrogen atom. $\Lambda$ is the cooling term, which includes both
bremsstrahlung and cyclotron radiation (Wu 1994), $G$ is the constant of
gravitation and $M$ is the mass of the white dwarf. 
We have included in Appendix A the derivation of the form of the $\Lambda$ 
term 
(used but not shown in Wu 1994) for the reader to assess the treatment of the 
cyclotron cooling. Note that we limit ourselves strictly to a 1-d flow: 
we do not for this
work consider spherically symmetric accretion or accretion in a dipolar
field geometry, as the form of (1) is particularly useful for our purposes.

The temperature and density structure of the postshock flow can be computed
from the coupled pair of nonlinear first order ODEs derived in Appendix
A. Rapid increases in the density at base of the flow make it more advantageous
to integrate using the velocity $v$ 
as the independent variable: (A9) should therefore be
inverted and then the dependence for the composite variable $\xi$ (see Appendix
A) can be written in terms of the
product of (A8) with the inverse of (A9). We therefore have
\[
\frac{dx}{dv}=
\]
\begin{equation}
\frac{\gamma(\xi-v)-v}{
-(\gamma-1)\frac{AC}{v^2}\sqrt{v(\xi-v)}\left(1+
  \epsilon_s\frac{4^{\alpha+\beta}}{3^{\alpha}}\frac{(\xi-
v)^{\alpha}v^{\beta}}{v_{a}^{\alpha+\beta}}\right)
+\frac{GM}{x^2}}
\end{equation}
and
\begin{equation}
\frac{d\xi}{dv}=\frac{d\xi}{dx}\frac{dx}{dv}=-\frac{GM}{x^{2}v}\frac{dx}{dv}.
\end{equation}
Here $-C$ is the mass transfer rate, $v_a=4v_s$ is the free-fall velocity at
the height of the shock, A is the coefficient for bremsstrahlung radiation,
$\alpha=3.85$ and $\beta=2$ are the coefficients corresponding to cyclotron
radiation in Wu, Chanmugam \& Shaviv (1994) and $\epsilon_s$ is the ratio of
the bremsstrahlung to cyclotron cooling time at the shock, as in Wu (1994).

Such a coupled system requires two boundary values.  Unfortunately this is not
an initial value problem: the initial value $x=\Rwd$ at $v=0$ is known, but the
variable $\xi$ is not -- it is constrained at the shock front by the
requirement that the pressure immediately behind the shock $P_s =
3Cv_a/4$. Moreover, this is a floating boundary condition as the freefall
velocity $v_a$ at the height of the shock is itself a result of the
computation.

We have also combined the coupled pair (4 and 5) into a single second
order nonlinear ODE in $x(v)$ (A14). Formally this now becomes an initial value
problem as at $x=\Rwd$, $v=0$ (as above) and $\frac{dv}{dx}=\infty$. However
because of poles in (A14) neither of these are accessible. Stepsizes must be
chosen appropriately in order to maintain the stability of the integration and
this route therefore appears to provide no computational advantage at this 
stage.

\section{Implementation}

We have used a Runge-Kutta-Merson method in a shooting and matching technique
implemented in the NAG D02HBF routine (Numerical Algorithms Group 1995) for
floating boundary conditions to solve the coupled pair of equations 4 and
5. Initial guesses for the parameters of the routine are derived from the
zero-gravity case. Convergence is generally reached within a few iterations,
thus permitting a rapid calculation of the postshock structure. The additional
computation time overhead by comparison with the closed integral form in Wu
(1994) is negligible.

We show in Figure 1 the temperature and density profile of the postshock flow
for the zero gravity case and our revised case including the
gravitational potential. We assume a 1\Msun white dwarf accreting
at 1 g s$^{-1}$ cm$^{2}$, corresponding to an accretion rate of $\sim 4\times
10^{15}$ g s$^{-1}$ over a fraction $f=0.001$ of the white dwarf, typical for
mCVs. The 0 MG, zero gravity case is the standard Aizu profile, while the 30 MG
zero gravity case is the same as the Wu (1994) profile.

\begin{figure}
\begin{center}
\epsfxsize=9.0cm 
\epsfbox{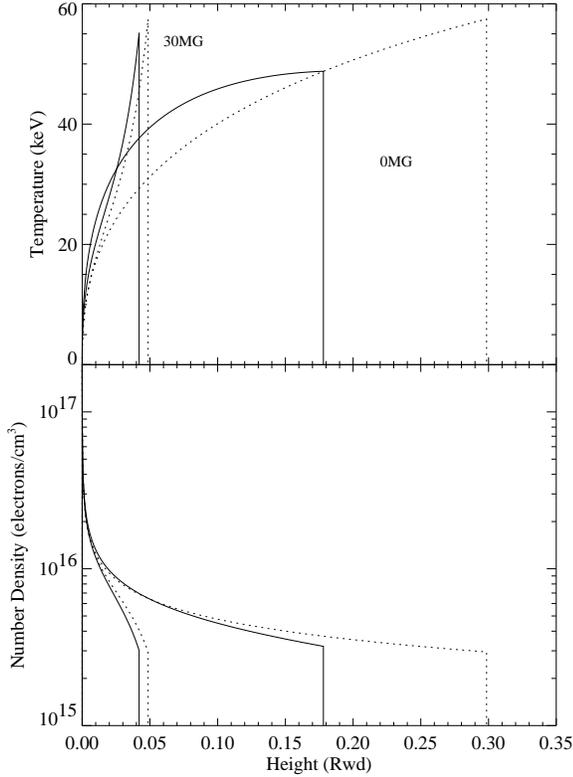}
\end{center}
\caption{Upper plot: The temperature profile for the case including the
gravitational potential (solid line) compared to the standard zero gravity
profile (dotted line) when cyclotron cooling is included (30 MG) and 
negligible
(0 MG). The vertical lines are at the position of the shock front. Lower plot:
as above but for the electron number density profiles. See text for details.}
\end{figure}

The effect of treating the gravitational potential more appropriately is
significant, especially in the case of the pure bremsstrahlung cooling (0 MG
field). The shock temperature is lower than that calculated from the Aizu
profile, because the pre-shock velocities at this height are less
than those at the surface of the white dwarf. However, there is significant
heating of the postshock flow itself through the release of gravitational
potential energy, so the temperature in the subsequent flow is higher than that
calculated from the Aizu profile. This leads to a flatter profile. In the case
where we include cyclotron cooling from a 30 MG field, the correction is not so
significant, as the height of the shock is in any case lower.

In Figure 2 we show the 
spectra for the pure bremsstrahlung case. The ratio plot
(lower panel) indicates that the new profile produces a harder spectrum with
fewer lines, at least for the 0.1 to 10 keV range shown here. This means that
fits to X-ray data using the new profile will, somewhat counterintuitively,
produce masses for the white dwarf which are lower than those determined using
the Aizu (1973) or Wu (1994) profiles (Cropper et al. 1998).

At higher energies, the Aizu spectrum becomes harder, as the postshock flow
immediately behind the shock is at a higher temperature. However, because for
1\Msun white dwarfs the shock temperature is $\sim 50$ keV, this is somewhat
higher than the typical range sampled by typical datasets available for these 
systems at present.

\begin{figure}
\begin{center}
\epsfxsize=9.0cm 
\epsfbox{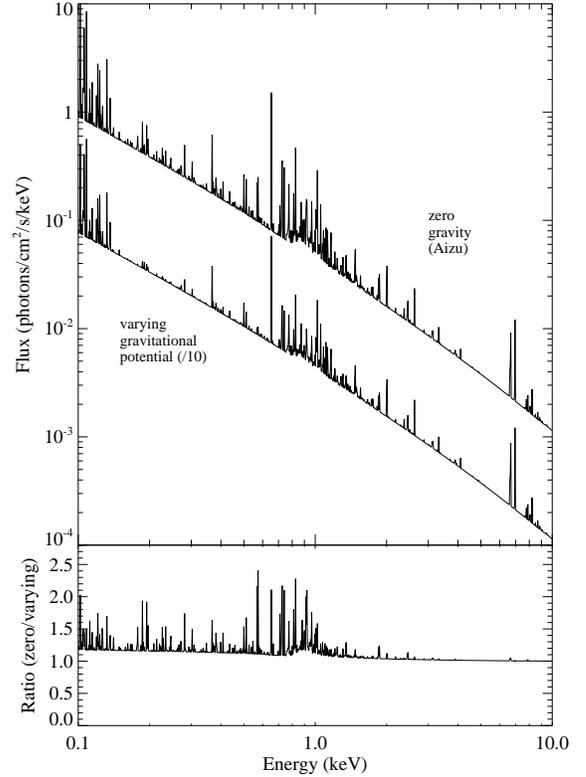}
\end{center}
\caption{Upper panel: The photon spectrum from the Aizu profile (above), and,
displaced downwards by a factor of 10 for clarity, that for the case when the 
effects of gravity are included. The parameters in Figure 1 are assumed
here, and the system distance is 100 pc. Lower plot: the ratio of the Aizu
spectrum to the spectrum from the varying gravitational potential.}
\end{figure}

\section{Improved fits to data}

Cropper et al. (1998) and Ramsay et al. (1998) derived masses for
the accreting magnetic white dwarfs in the {\it Ginga} archives. Some of these exceeded
1.0 \Msun. This indicates that the shock height in these particular systems is
likely to be significant, especially for the IPs in the sample. Therefore their
white dwarf mass determinations can be improved significantly by taking into
account the revised temperature and density profiles calculated above.

\begin{table*}
\begin{tabular}{lllll}
\hline
System&$\mu$=0.5+IA&$\mu$=0.615+IA&$\mu$=0.615+G+IA&$\mu$=0.615+G+PC\\
      &$M_{\odot} (\chi^{2}_{red})$&$M_{\odot}
      (\chi^{2}_{red})$&$M_{\odot} (\chi^{2}_{red})$&$M_{\odot} (\chi^{2}_{red})$\\
\hline
AM Her&1.22 (1.43: 1.15--1.30) & 1.12 (1.28: 1.06--1.20) & 1.06 
(1.35: 0.97--1.16) & 0.85 (0.61: 0.76--0.90)\\
EF Eri&0.88 (1.00: 0.82--0.94) & 0.81 (1.07: 0.76--1.15) & 0.80 (1.02:
0.75--0.85)& 0.80 (0.91: 0.68--1.17) \\
BY Cam&1.08 (1.53: 1.00--1.25) & 1.27 (1.43: $>$0.63) & 1.18 (1.38:
0.76--1.27)& 0.98 (1.16: $>$0.65)\\
V834 Cen&0.5 (0.88: 0.25--1.00) & 0.5 (0.82: $<$1.3) & 0.64 (0.89:
$<$1.15)& 0.54 (0.95: $<$1.25)\\
QQ Vul&1.13 (1.40: 0.95--1.30) & 1.30 (1.32: $>$0.7) & 0.95 (1.27:
$>$0.6)&0.60 (1.18: 0.4--1.4)\\
\hline
EX Hya& 0.52 (0.65: 0.42--0.60)& 0.45 (0.59: 0.32--0.52)& 0.50
(0.59: 0.44--0.56)& 0.46 (0.56: 0.42--0.50)\\
AO Psc& 0.40 (0.93: $<$0.80) & 0.45 (0.90: $<$0.62) & 0.36 (0.89:
$<$0.65)& 0.56 (0.94: 0.36--0.72) \\
FO Aqr& 1.22 (0.83: $>$1.00)& 1.12 (0.77: $>$0.89) & 1.07 (0.73:
0.91--1.22)& 0.92 (0.73: 0.58--1.22)\\
TV Col& 1.20 (0.84: $>$0.95) & 1.22 (0.93: 1.03--1.36) & 1.21 (0.82:
$>$1.06) & 1.3 (1.21: $>$0.9)\\
BG CMi& 1.25 (0.62: 1.05--1.39) & 1.36 (0.93: $>$1.18) & 1.19 (0.95:
$>$0.95) &1.09 (0.58: $>$0.94) \\
TX Col& 0.55 (1.34: 0.44-0.64)& 0.48 (1.28: 0.41--0.54) & 0.48 (1.23: 0.37--0.53) & 0.48
(1.26: 0.39--0.54)\\
PQ Gem& 1.35 (1.08: $>$ 1.05)& 1.32 (1.04: $>$1.15)& 1.21 (1.06:
$>$1.08) & 1.29 (1.04: $>$1.17)\\
AE Aqr& 0.3 (0.76: $<$1.15)& 0.62 (0.68: 0.4--1.15)& 0.6 (0.69:
$<$1.0) & 0.6 (0.66: $<$1.0) \\
\hline
\end{tabular}
\caption{The best fit to the mass of the white dwarf in the magnetic
CVs which were observed using {\sl Ginga} using variants of the
stratified accretion column model of Cropper et al. (1998). The
upper panel shows the results of the synchronous (or very near)
systems (the Polars), while the lower panel shows the results of the
asynchronous systems (the Intermediate Polars).  The column with the
mean molecular mass of the plasma, $\mu$=0.5, corresponds to that
assumed in Cropper et al. (1998). (Some of these new values
differ marginally to those quoted there since in this paper the
accretion shock has been divided into 100 vertical elements as opposed to 50
there). The remaining columns assume $\mu$=0.615 with the last 2
columns including the gravitational term described in this paper (+G).
IA refers to an ionised absorber while PC refers to partial covering
of cold absorber. The range in mass is the 90 per cent confidence
interval. In all cases a cold absorber (for interstellar absorption) 
and a fluoresence line fixed at 6.4keV have been included.}
\end{table*}

\begin{table*}
\begin{center}
\begin{tabular}{llll}
\hline
Satellite&$\mu=0.5$+IA&$\mu$=0.615+IA&$\mu$=0.615+IA+G\\
         &$M_{\odot}$ (\rchi)&$M_{\odot}$ (\rchi)&$M_{\odot}$ (\rchi)\\
\hline
{\sl RXTE}& & & \\
(1)& 1.03 (1.04: 0.94--1.11)& 0.98 (1.04: 0.85--1.03) & 0.89 (1.04: 
0.84--1.00)\\
(2)& 0.89 (0.76: 0.83--1.02)& 0.83 (0.76: 0.75--0.92) & 0.82 (0.76: 
0.74--0.91)\\
\hline
{\sl ASCA}& & & \\
(GIS2)&1.05 (0.88: 0.81--1.32)& 1.01 (0.88: 0.75--1.25) & 0.96 (0.89: 
0.72--1.16)\\
(GIS3)&1.38 (1.06: $>$1.18)& 1.37 (1.07: $>$1.12) & 1.25 (1.08: 1.05--1.36)\\
(GIS2\&3)& 1.25 (0.97: 1.08--1.38)& 1.23 (0.97: 0.98--1.36) & 1.15 
(0.98: 0.96--1.25)\\
(SIS0)& 1.26 (1.05: $>$0.96)& 1.26 (1.05: $>$0.86) & 0.98 (1.07: 0.78--1.32)\\
(SIS1)&1.40 (1.11: $>$1.27)& 1.40 (1.10: $>$1.23) & 1.31 (1.10: $>$1.07)\\
(SIS0\&1)&1.40 (1.05: $>$1.24)& 1.40 (1.05: $>$1.17) & 1.19 (1.08: 
1.00--1.36)\\
(GIS\&SIS)&1.33 (1.03: $>$1.22)& 1.31 (1.03: $>$1.13) & 1.15 (1.05: 
1.02--1.27)\\
\hline
{\sl Ginga}& & &\\
 &1.28 (0.46: $>$1.17)& 1.23 (0.46: 1.08--1.38)& 1.19 (1.19: 1.06--1.32)\\
\hline
Hellier (1)& &0.91--1.29 &\\
Hellier (2)& &0.74--1.14& \\
Refining& 1 \& 2 & 0.78--1.03& \\
\hline
\end{tabular}
\end{center}
\caption{A comparision of mass estimates using XY Ari data using the
models described in Table 1. No Iron fluorescence component at 6.4keV
has been added. The range in mass is for the 90 per cent confidence
level. The data in the first column have previously been reported in
Ramsay et al. (1998).}
\end{table*}

We have proceeded as in Cropper et al. (1998) and Ramsay et al.
(1998), so that a direct comparison can be made with the earlier mass
derivations. The results are shown in Table 1 for the former paper and
in Table 2 for the latter. An intermediate level of refinement is also
shown in the tables: the earlier papers assumed a pure Hydrogen plasma
($\mu=0.5$) for calculating 
the temperature and density profiles of (but not the
spectra from) the postshock flow. Tables 1 and 2 therefore show the
original mass determinations, the revised determinations with a cosmic
plasma where $\mu=0.615$, and then the present calculations including
the gravitational potential and $\mu=0.615$. 

In addition to cold interstellar absorption, Cropper et al. (1998) and
Ramsay et al. (1998) used a partially ionised absorber to account for
the absorption within the system of the preshock flow.  In the
presence of density inhomogeneities in the accretion flow, and
arc-shaped accretion footprints, the treatment of the absorption is a
complex matter (Done \& Magdziarz 1998, Rainger 1998, in preparation).
The final ingredients for the absorption model would include ionised
absorption for the finely divided flow and a range of partial covering
fractions for the density inhomogeneities, all integrated over a range
of pathlengths dependent on the viewing angle to the flow. As noted
earlier, this is beyond the scope of this work. However we have
included in Table 1 a final column which shows the fits
including the gravitational potential and $\mu=0.615$ but now assuming
a partial covering of cold material, rather than the partially ionised
absorber. This may be useful for comparison with other workers, as
the partially ionised absorber model is not generally available.

As expected from Figures 1 and 2, the results in Tables 1 and 2
indicate reduced masses for the massive white dwarfs in these systems
when the gravitational potential is included in the shock structure.
In the case of Table 1 the average (unweighted) 
reduction in mass of those systems
above 1.0 \Msun in Cropper et al. (1998) is approximately
0.1 \Msun. For those below 1.0 \Msun the masses are slightly ($\sim
0.03$\Msun) increased.  There is little difference between the masses
assuming $\mu=0.5$ and $\mu=0.615$ in the zero gravity case. There is
some scatter in the trends because the fitting routine optimises parameters
such as mass transfer rate in addition to the white dwarf mass, and in
some cases new global optima are found. The masses derived using the
partial fraction covering model for the absorption are slightly lower
still: for those systems previously above 1.0 \Msun the average
reduction in mass is a further $\sim0.12$ \Msun, while that for the
less massive systems is unchanged.

In the case of Table 2, we also see reductions $\sim0.1$ \Msun as the effect of
the gravitational potential is included in the mass determinations for XY
Ari. 

\section{Conclusions}

The results in Tables 1 and 2 show a reduction in the mean of the 
derived masses for the white dwarf when the effects of gravity are included. 
The reductions are greatest for more massive white dwarfs,
low specific accretion rates and low magnetic fields, because the
shock height is greater under these conditions. The determinations for
lower mass white dwarfs remain largely unaffected.

Comparing the revised determinations with those from other methods, we
note that agreement in the case of XY Ari is significantly improved:
for the {\it RXTE} data the best mass and $2\sigma$ errors are in
close agreement with those determined from the eclipse duration in
Ramsay et al. (1998). The {\it Ginga} and {\it Asca} determinations
remain significantly higher, however: Ramsay et al. (1998) explored some of the
possible explanations for this. In the case of the systems in Table
1 for which other determinations exist (see also Cropper et al. 1998) we
note close agreement in the case of the lower mass systems EX Hya as
determined from the X-ray line data (Fujimoto \& Ishida 1997, Ishida
1999, Ezuka \& Ishida 1999), and with 
AO Psc and TX Col (Hellier et al. 1996, Ezuka \& Ishida 1999). 
In the case of FO Aqr, where the mass is higher at $\sim 1$ \Msun, there is
also close agreement with that determined from {\it Asca} data (Ezuka \& Ishida
1999). The {\it Ginga} data for
AE Aqr and QQ Vul are relatively poorer, and do not constrain the
masses strongly: both are nevertheless consistent with those
determined elsewhere (Welsh, Horne \& Gomer 1994, Mukai \& Charles
1987). Only in the case of AM Her is the mass we derive still
significantly higher than those from other techniques (G\"{a}nsicke et
al. 1998, and to a lesser extent, Mukai \& Charles 1987). 
However the effect of the absorption
model is particularly strong, so that if a partial covering model is
assumed, even here the mass of 0.85 \Msun is not controversially
high (Mukai \& Charles 1987).

We note that any technique for determining the mass of the white dwarf
will have both systematic biases built into the method and
uncertainties arising from poor constraints in some of the input
parameters, such as inclination for the kinematically derived masses,
or perhaps instrumental effects as in the case of XY Ari (Ramsay et
al. 1998). 

Our results indicate the importance of the absorption model on the
mass determinations. The effect can be minimised by excluding the
softest photons from the fits to the data, especially in the case of
CCD data with sufficient spectral resolution. The resulting models can
be used retrospectively to isolate the absorption at lower energies.
Clearly, with shock temperatures in excess of 40 keV for 1 \Msun white
dwarfs, X-ray spectra with sufficient data quality extending to
approximately equivalent energies are beneficial for the extraction of
mass information.

\section{ACKNOWLEDGEMENTS}

We are very grateful to Lee McDonald and Mitchell Berger for their help, and
to Andy Beardmore who pointed out the need to use cosmic abundances to
determine the shock temperature appropriately. KW
acknowledges the support from the Australian Research Council through an
an Australian Research Fellowship.

\appendix

\section{Hydrodynamic Equations}

The time dependent mass continuity, momentum and energy equations are 
(see e.g.\ Frank, King \& Raine 1992):
\begin{equation}
\frac{\partial \rho}{\partial t}+ \nabla \cdot(\rho {\bf v}) = 0
\end{equation}
\begin{equation}
\rho \frac{\partial {\bf v}}{\partial t} + 
          \rho{\bf v}\cdot \nabla{\bf v} = -  \nabla {\bf P}+{\bf f}
\end{equation}
\begin{eqnarray}
\lefteqn{
\frac{\partial}{\partial t}(\frac{1}{2}\rho v^2+\rho\varepsilon)+
{\bf \nabla}\cdot[(\frac{1}{2}\rho v^2+\rho\varepsilon+P){\bf v}] 
                     = } \hspace{3cm} \nonumber \\
 & &{\bf f}\cdot {\bf v} - \nabla \cdot {\bf F}_{\rm rad}\ .
\end{eqnarray}   
Here we ignore heat conduction and fluid viscosity.  

\subsection{The Radiative Cooling}

By Gauss' Theorem, we can replace the volume integral of the divergence 
of the radiation flux $\nabla \cdot {\bf F}_{\rm rad}$ in (A3) by a 
surface integral   
\[
\int d^3x\ \nabla \cdot {\bf F}_{\rm rad}\ 
   = \oint d{\bf s}\ \cdot{\bf F}_{\rm rad}\  
\]  

Suppose the radiative flux ${\bf F}_{\rm rad}$ consists of  
of two components: an optically thin bremsstrahlung term ${\bf F}_{\rm br}$ 
and an optically thick cyclotron term ${\bf F}_{\rm cyc}$, thus,\  
\[
{\bf F}_{\rm rad} = {\bf F}_{\rm br} + {\bf F}_{\rm cyc}\ . 
\]
Since bremsstrahlung emission is optically thin and isotropic, the surface 
integral of its flux at any arbitrary closed surface $S'$ is 4$\pi$ times 
the volume integral of the bremsstrahlung emissivity in a volume $V'$
enclosed by the surface. Therefore, we have  
\[
\oint_{_{S'}} d{\bf s}\cdot {\bf F}_{\rm br}\ 
  = \int_{_{V'}} d^3x\ (4\pi j_{\rm br})\ = 
  \int_{_{V'}} d^3x\ \Lambda_{\rm br}\ , 
\]
where $\Lambda_{\rm br}$ is the energy loss per unit volume due to  
bremsstrahlung emission, which is the bremsstrahlung cooling function. Since 
the volume $V'$ is arbitrarily chosen, we have in all space  
\[
\nabla \cdot {\bf F}_{\rm br} = \Lambda_{\rm br}
    = A\rho^{2}\sqrt{\frac{P}{\rho}}\ , 
\]
where $A$ is the constant for bremsstrahlung emission, $P$ is the pressure and 
$\rho$ is the density.

%

To evaluate the volume integral of the divergence of the cyclotron flux, 
we consider the postshock emission region as a cylinder with its symmetry 
axis parallel to the magnetic field. Because cyclotron emission is not 
optically thin and it is not isotropic, we cannot equate 
$\nabla \cdot {\bf F}_{\rm cyc}/4 \pi $ to the 
cyclotron emissivity $j_{\rm cyc}$. As the volume integral of 
the divergence of the flux is the same as the integration of the flux 
over the surface enclosing the volume, we need only to evaluate the 
flux at the surface of the cylinder. Cyclotron emission 
is strongly beamed such that most power propagates preferentially in the 
direction perpendicular to the magnetic field, the vector products of 
$d{\bf s}\cdot {\bf F}_{\rm cyc}$ at the top surface $S_1$ and the 
bottom surface $S_3$ of the cylinder are small in comparison with that 
at the side surface $S_2$ (this is not equivalent to assuming that the 
fluxes at the surfaces $S_1$ and $S_3$ are small). As an approximation we  
neglect their contribution to the surface integral. Then we have   
\[
\int_{_{V_{\rm c}}} d^3x\ \nabla \cdot {\bf F}_{\rm cyc}\ 
   \simeq \int_{_{S_2}} d{\bf s}\ \cdot{\bf F}_{\rm cyc}\  ,  
\]   
where $V_{\rm c}$ is the total volume of the cylinder. 

Suppose the spectrum of the cyclotron emission is optically thick up 
to a frequency $\nu_{*} (= \omega_*/2 \pi)$ after which it is optically 
thin. For parameters typical of the accretion shocks in mCVs the 
optically thin 
cyclotron intensity falls rapidly with frequency as $\sim\nu^{-8}$ 
(see Chanmugam et\, al. 1989). As a first approximation we can 
neglect the contribution of the optically thin cyclotron emission to 
the total cyclotron cooling process determining the structure of 
the postshock accretion flow. Thus
\[
\int_{_{S_2}}d{\bf s} \cdot {\bf F}_{\rm cyc} \simeq 
       \pi^2 D \int_{R_{_{\rm WD}}}^{x_{\rm s}} dx
          \int_{0}^{\nu_{*}} d\nu  B_{_{\rm RJ}}(\nu)  
\]
where $B_{_{\rm RJ}} = 2kT\nu^{2}/c^{2}$ is the Rayleigh-Jeans intensity, 
and $x_{\rm s}$ is the shock height and $D$ is diameter of the cylinder 
(which is the accretion column) respectively. 

If we divide the accretion column in layers of height $dx$, the 
cyclotron luminosity, which is the total energy loss due to cyclotron 
emission, is the sum of the contribution of these layers:  
\[
L_{\rm cyc} = 
 \int_{R_{_{\rm WD}}}^{x_{\rm s}} dx\ \frac{dL_{\rm cyc}}{dx}\ . 
\]
Thus, we obtain an effective local cyclotron cooling term  
\[
\Lambda_{\rm cyc}  = \frac{dL_{\rm cyc}}{dx} = 
                     \pi D \frac{kT\omega_{*}^3}{12\pi^2 c^2}\ .
\] 

Now from Wada et al.\ (1980)
\[
\omega_{*}(x) \simeq 9.87 \omega_{\rm c} \left[ \frac{\Theta}{10^{4}} 
\right]^{0.05} 
\left[ \frac{T}{10^8 {\rm K}} \right]^{0.5}\ , 
\]
where the cyclotron frequency $\omega_{\rm c} = eB/m_{\rm e}c$ and 
the dimensionless plasma parameter $\Theta = 2\pi n_{\rm e}(x)D/B$ 
(normally written $\Lambda$). Here $B$ is the magnetic field and 
$n_{\rm e}$ is the electron mumber density. For constant $B$,
\[
\Lambda_{\rm cyc} \propto \omega_{*}^{3}T 
   \propto n_{\rm e}^{0.15}T^{2.5}
\]

We define two quantities
\[
 t_{\rm cyc} \equiv 
       \frac{3}{2} (n_e + n_i) kT \frac{1}{\Lambda_{\rm cyc}}
\]
and
\[
 t_{\rm br} \equiv 
      \frac{3}{2} (n_e + n_p) kT \frac{1}{\Lambda_{\rm br}}   
\]
for the cyclotron and bremsstrahlung cooling respectively ($n_i$ is the ion 
number density). 
In terms of these two quantities, the total effective cooling term is 
\[
\Lambda = \Lambda_{\rm br} + \Lambda_{\rm cyc} = 
    \Lambda_{\rm br}\left( 1+\frac{t_{\rm br}}{t_{\rm cyc}}\right)
\]
with  
\[
\frac{t_{\rm br}}{t_{\rm cyc}} \propto \frac{T^{2}}{n_{\rm e}^{1.85}}\ .
\]
We assign a proportionality $\epsilon(x) = t_{\rm br}/t_{\rm cyc}$ 
and scale it in terms of the pressures and densities at the shock, $\epsilon_{s}$. With the ideal gas law, we eliminate the temperature and obtain  
\[
  \epsilon(x) = \left.\frac{t_{\rm br}}{t_{\rm cyc}}\right|_{x_{\rm s}} 
\left(\frac{P}{P_{\rm s}}\right)^{2.0}
  \left(\frac{\rho_{\rm s}}{\rho}\right)^{3.85}.
\]
Thus, we arrive at the effective cooling term determining the dynamics of 
the flow as that used in Wu (1994)
\begin{equation}
\Lambda = \Lambda_{br}\left[1+\epsilon_{\rm s}
  \left(\frac{P}{P_{\rm s}}\right)^{2.0}\left(\frac{\rho_{\rm s}}{\rho}
  \right)^{3.85}\right]\ .
\end{equation}

\subsection{Steady State 1-D Hydrodynamic Equations including Gravity}

For a stationary state, the time-derivative in the hydrodyanamic equations 
are zero. As we consider a 1-dimensional flow channelled along a cyclinder 
parallel to the magnetic field, (A1) to (A3) can be reduced to 
\begin{equation}
\frac{d}{dx}(\rho v)= 0
\end{equation}
\begin{equation}
\rho v \frac{dv}{dx}+\frac{dP}{dx}= f
\end{equation}
\begin{equation}
\frac{d}{dx}[(\frac{1}{2}\rho v^2+\rho\varepsilon+P)v]=vf-\Lambda.
\end{equation}
The symbols are as defined in Section 2, and $f$ is the force term,
$\varepsilon$ is the internal energy of the gas.
As the cross section of the accretion column is small in comparison with 
the white dwarf radius, it is adequate to let  
\[
f = \rho g = -\frac{\rho GM}{x^2}\ . 
\]
Noting from (A5) that
\[
\frac{d}{dx}(\rho v^{2})=v\frac{d}{dx}(\rho v) + \rho v\frac{dv}{dx}
                        =\rho v\frac{dv}{dx}    
\]
we obtain for (A6)
\begin{equation}
\frac{d}{dx}(\rho v^2+P) = -\rho\frac{GM}{x^2}\ .
\end{equation}
Substituting for $f$ from (A6) in (A7) yields 
\[
v\frac{d}{dx}(\rho\varepsilon-\frac{1}{2}\rho v^2)+
(\frac{1}{2}\rho v^2+\rho\varepsilon+P)\frac{dv}{dx} = -\Lambda.
\]
Using $\varepsilon=\frac{P}{\rho}\frac{1}{\gamma-1}$, and again (A5), we
obtain
\[
\frac{v}{\gamma-1}\frac{dP}{dx}+\frac{\gamma P}{\gamma-1}\frac{dv}{dx}=
 -\Lambda
\]
and thus
\begin{equation}
v\frac{dP}{dx}+\gamma P\frac{dv}{dx} = -(\gamma-1)\Lambda
\end{equation}
and in (A5), (A8) and (A9) we recover the hydrodynamic equations of Wu (1994),
but with the additional gravitational force term in the momentum equation.


\subsection{Coupled First Order Form}

We now substitute the variable $\xi=v+(P/\rho v)$ in place of $P$ in the
hydrodynamic equations. Thus the momentum equation (A8) becomes
\[
\frac{d}{dx}(\rho v \xi) = \frac{-GM\rho}{x^2}.
\]
Using (1) we obtain
\[
\rho v\frac{d\xi}{dx} = \frac{-GM\rho}{x^2}
\]
and thus
\begin{equation}
\frac{d\xi}{dx} = \frac{-GM}{x^{2}v}.
\end{equation}

Substituting for $\xi$ in the energy equation (A7) and noting from (A1) that
$\rho v = $constant$ = C$ (where $-C$ is the mass transfer rate 
  $\Mdot$),   
\[
vC\frac{d\xi}{dx} - vC\frac{dv}{dx} + \gamma C(\xi-v)\frac{dv}{dx} = 
                  -(\gamma-1)\Lambda \ . 
\]
Thus
\begin{equation}
\frac{dv}{dx}=\frac{-\frac{\gamma-1}{C}\Lambda+\frac{GM}{x^2}}
                   {\gamma(\xi-v)-v}
\end{equation}

%

Now substituting for $\Lambda_{\rm br}$ in (A4) we have
\[
\Lambda = A\rho^{2}\sqrt{\frac{P}{\rho}}\left[
           1+\epsilon_{\rm s}\left(\frac{P}{P_{\rm s}}\right)^{\alpha}
                         \left(\frac{\rho_{\rm s}}{\rho}\right)^{\beta} \right]
\]
thus
\begin{equation}
\Lambda = \frac{AC^2}{v^2}\sqrt{v(\xi-v)}\left[
          1+\epsilon_{s}\frac{4^{\alpha+\beta}}{3^{\alpha}}
              \frac{(\xi-v)^{\alpha}v^{\beta}}{v_a^{\alpha+\beta}}\right]
\end{equation}
where we have used zero pre-shock pressure (so that across the shock from (2)
$\rho_{\rm a} v_{\rm a}^2 = \rho_{\rm s} v_{\rm s}^2 + P_{\rm s}$ and 
$\rho_{\rm a} v_{\rm a} = \rho_{\rm s} v_{\rm s}$) to eliminate
$P_{\rm s}$ and $\rho_{\rm s}$ and symbols are defined in Section 2. 
Combining (A10) to (A12) we obtain equations (4) and (5) in Section 2. 


\subsection{Second Order Form}

We now proceed to the 2nd order form in $v$. 
We eliminate $\xi$ by substituting further $w=(\xi/v)-1$ and 
\[
\epsilon_{\rm s}^{\prime} = \epsilon_{\rm s}\frac{4^{\alpha+\beta}}{3^{\alpha}}
              \frac{1}{v_{\rm a}^{\alpha+\beta}}.
\]
Equation (4) becomes
\begin{equation}
\frac{dx}{dv} = \frac{v(\gamma w - 1)}{-(\gamma-1)\frac{AC}{v}\sqrt{w}\left( 
1+\epsilon_{\rm s}^{\prime} v^{5.85}w^{2}\right)+\frac{GM}{x^2}}.
\end{equation}
For $\epsilon_{\rm s}^{\prime} = 0$ (no cyclotron) this is a quadratic in
$\sqrt{w}$ and two roots can be found analytically by rearranging (A13) and
applying the standard formula for quadratic roots. Otherwise this is a quintic
with no general analytic form for the roots, which then have to be found
numerically for the particular case.

Differentiating (A13) again we have
\begin{eqnarray}
\frac{d^{2}x}{dv^2} & = &  \left\{ \left[-(\gamma-1)\frac{AC\sqrt{w}}{v^2}+
    \frac{GM}{x^{2}v}\right]\gamma\frac{dw}{dv} - \right.  \nonumber \\
& & (\gamma w-1)\left[
-(\gamma-1)\frac{AC}{v^2}\left(\frac{1}{2\sqrt{w}}\frac{dw}{dv}-
 \frac{2\sqrt{w}}{v}\right)- \right. \nonumber \\
& & \hspace*{2.5cm} \left.\left. 
\frac{GM}{x^{2}v}\left(\frac{2}{x}\frac{dx}{dv}+
 \frac{1}{v}\right)\right]\right\} \nonumber \\
& & \overline{\hspace*{1cm}\left[-(\gamma-1)\frac{AC\sqrt{w}}{v^2}+
 \frac{GM}{x^{2}v}\right]^2\hspace*{1cm}} 
\end{eqnarray}
where, from (5)
\begin{equation}
\frac{dw}{dv}=-\frac{GM}{x^{2}v^{2}}\frac{dx}{dv}-\frac{w+1}{v}.
\end{equation}
Thus by substituting for $\sqrt{w}$ from the roots of (A13), and
$\frac{dw}{dv}$ from (A15) into (A14) we obtain a 2nd order nonlinear ODE in
$x$ and $v$ alone.

\end{document}